\documentclass[aps,prd,nofootinbib,preprint,floatfix,unsortedaddress]{revtex4-1}
\usepackage{setspace}
\usepackage{graphicx}
\usepackage{dcolumn}
\usepackage{multirow}
\usepackage{subfigure}
\usepackage{times,mathptm}
\usepackage{float}
\usepackage{color}
\usepackage{amsmath,amsfonts}
\usepackage{mathptmx}
\usepackage{mathrsfs}
\usepackage{bbm}
\usepackage{bm}
\usepackage{xfrac}

\newcommand{\bea}{\begin{eqnarray}}
\newcommand{\eea}{\end{eqnarray}}

\renewcommand{\d}{\delta}

\newcommand{\ihat}{\boldsymbol{\hat{\textbf{\i}}}}

\newcommand{\tQ}{\widetilde{Q}}

\renewcommand{\b}{\beta}
\renewcommand{\a}{\alpha}

\newcommand{\tr}{\text{Tr}}

\newcommand{\tP}{\widetilde{P}}

\newcommand{\vx}{{\vec{x}}}
\newcommand{\vy}{{\vec{y}}}

\newcommand{\vk}{{\vec{k}}}

\newcommand{\m}{\mu}

\newcommand{\D}{\Delta}

\newcommand{\vq}{\vec{q}}

\newcommand{\oh}{\frac{1}{2}}

\newcommand{\non}{\nonumber}

\newcommand{\rf}[1]{(\ref{#1})}
\newcommand{\ra}{\rightarrow}
\newcommand{\pa}{\partial}
\renewcommand{\vec}[1]{\bm #1}

\usepackage{ulem}

\begin{document}

\title{Effective Polyakov line action from strong lattice couplings
to the deconfinement transition} 
 
\author{Jeff Greensite}
\affiliation{\singlespacing Physics and Astronomy Department, \\ San Francisco State
University, San Francisco, CA~94132, USA}

\author{Kurt Langfeld}
\affiliation{\singlespacing School of Computing \& Mathematics, University of Plymouth, Plymouth, PL4 8AA, UK}
\date{\today}
\vspace{60pt}
\begin{abstract}

\singlespacing
 
      We calculate the effective Polyakov line action corresponding to SU(2) lattice gauge theory 
on a ${16^3 \times 4}$ lattice via the ``relative weights'' method.  We consider a variety of lattice couplings, ranging from 
$\b=1.2$ in the strong-coupling domain, to $\b=2.3$ at the deconfinement transition, in order to study how the effective action evolves with $\b$.  Comparison of Polyakov line correlators
computed in the effective theory and the underlying gauge theory is used to test the validity of the effective action for $\b > 1.4$,
while for $\b=1.2, 1.4$ we can compare our effective action to the one obtained from a low-order strong-coupling expansion.  Very good agreement is found at all couplings.  We find that the effective action is given by a simple expression bilinear in the Polyakov lines. The range of the bilinear term, away from strong coupling, grows rapidly in lattice units as $\b$ increases.         
\end{abstract}

\maketitle

\singlespacing
\section{\label{sec:intro}Introduction}
 
     In a recent article \cite{Greensite:2013yd}  we have applied a technique, which we call the ``relative weights'' method, to determine the effective Polyakov line action (PLA) corresponding to an SU(2) lattice gauge theory.  The effective Polyakov line action $S_P$ is defined as the action which results from integrating out all gauge and matter degrees of freedom in the lattice gauge theory, whose action is denoted $S_L$, under the constraint
that Polyakov line holonomies are held fixed.  In temporal gauge, where the timelike link variables are set to the identity
matrix except on a single timeslice at, say, $t=0$, we have~\footnote{For convenience, we adopt a sign convention for the action such that the Boltzman weight is proportional to $\exp[+S]$.}
\bea
\exp\Bigl[S_P[U_{\vx}]\Bigl] =    \int  DU_0(\vx,0) DU_k  D\phi ~ \left\{\prod_{\vx} \d[U_{\vx}-U_0(\vx,0)]  \right\}
 e^{S_L} \ .
\label{S_P}
\eea
where $\phi$ denotes any matter fields, scalar or fermionic, coupled to the gauge field.
  
     In pure SU(2) lattice gauge theory, which is the case we will consider in this article, $S_P$ can depend only on the trace
of Polyakov line holonomies.  Consider a Fourier expansion of the trace
\bea
                P_{\vx} \equiv \oh \tr[U_{\vx}] =  a_0 + \oh \sum_{\vq\ne 0} \Bigl\{ a_{\vq} \cos(\vq \cdot \vx) + 
                       b_{\vq} \sin(\vq \cdot \vx) \Bigr\} \ ,
\label{dft}
\eea
where the sum  runs over all wavevectors $\vq$ on a cubic lattice of volume $L^3$, and ${a_{\vq}=a_{-\vq}, ~ b_{\vq}=-b_{-\vq}}$    are real-valued.  The relative weights method allows us to compute the derivatives of the action with respect to
any of the Fourier components
\bea
             \left( {\pa S_P \over \pa a_{\vk}} \right)_{a_{\vk}=\a}  ~~,~~   \left( {\pa S_P \over \pa b_{\vk}} \right)_{b_{\vk}=\a}  \ ,
\label{grad1}
\eea 
and from these derivatives we are able to deduce the action itself. In ref.\ \cite{Greensite:2013yd}  we found that the effective action is a simple
bilinear expression in terms of the Polyakov lines:
\bea
           S_P =  \oh c_1 \sum_{\vx} P^2_{\vx} -  2c_2 \sum_{\vx \vy} P_{\vx} Q(\vx - \vy) P_{\vy} \ ,
\label{SP1}
\eea
where
\bea
           Q(\vx-\vy) = \left\{  \begin{array}{cc}
                     \Bigl(\sqrt{-\nabla_L^2}\Bigr)_{\vx \vy}  &  |\vx-\vy| \le r_{max} \cr
                       0 & |\vx-\vy| > r_{max} \end{array} \right. \ .
\label{Q}
\eea
and $\nabla_L^2$ is the lattice Laplacian.  Our method also determines the constants $c_1, c_2$ and the range $r_{max}$.  
The validity of this expression was tested by comparing Polyakov line correlators computed by numerical simulation of the effective theory based on $S_P$, and of the underlying lattice gauge theory.  We found accurate agreement, down to correlator magnitudes on the order of $10^{-5}$, between the two sets of correlators.

     A limitation of the work in \cite{Greensite:2013yd} is that it was carried out for a single gauge coupling $\b=2.2$ at $N_t=4$
lattice spacings in the time direction.  An obvious question is how the effective action evolves as $\b$ varies from strong couplings to weaker couplings, up to the deconfinement phase transition.  Answering this question will also allow us to check how the relative weights method performs over a range of couplings, rather than just a single coupling.  At very strong couplings
the effective action can be computed analytically, and only the nearest-neighbor term in the effective action is significant.  At next to leading order, we have~\footnote{For a higher-order computation, cf.\ \cite{Fromm:2011qi}.}
\bea
           S_P &=& \b_P   \sum_\vx \sum_{i=1}^3   P_\vx P_{\vx+ \ihat}  \ ,                
\non \\
     \b_P &=&  4 \left[1 + 4N_t \left({I_2(\b) \over I_1(\b)}\right)^4 \right]  \left({I_2(\b) \over I_1(\b)}\right)^{N_t} \ .
\label{Sp_strong}
\eea
where $N_t$ is the lattice extension in the periodic time direction.
In contrast, at $\b=2.2$, we found it necessary to include couplings among Polyakov lines separated by distances up to 
$r_{max}=3$ lattice units.  We would like to study how the action, and in particular the range of the bilinear term $r_{max}$, evolves from strong coupling through to the deconfinement transition.  

   In this article we extend our previous work by computing $S_P$, via the relative weights method, for a variety of lattice couplings, starting from a strong coupling value of $\b=1.2$, and proceeding to the deconfinement transition at $\b=2.3,~N_t=4$.  All our numerical computations are carried out on a $16^3 \times 4$ lattice volume for the SU(2) lattice gauge theory, while simulations of $S_P$ are carried on a three dimensional $16^3$ lattice volume.  Our main finding is that the effective action is well described by a bilinear form throughout the range of lattice couplings, and that the range of the kernel $Q(\vx-\vy)$, beyond the strong-coupling limit, increases rapidly in lattice units with increasing $\b$.
   
\section{\label{sec:rw} Procedure}

    The relative weights method allows us to calculate numerically the difference
\bea
           \D S_P = S_P[U'_\vx] - S_P[U''_\vx]
\eea
of Polyakov line actions, evaluated at configurations $U'_\vx$ and $U''_\vx$ respectively, 
which are nearby in the configuration space of
Polyakov line holonomies.  The method is based on the fact that, while it may be difficult to evaluate the integral in
\rf{S_P} directly, the ratio (or ``relative weights'') 
\bea
e^{\D S_P} = {\exp[S_P[U'_\vx]]\over \exp[S_P[U''_\vx]]}
\eea
can be expressed in a form
which is more amenable to numerical simulation.   Let $S'_L, S''_L$ represent the lattice action with timelike links $U_0(\vx,0)$
fixed to $U'_\vx$ and $U''_\vx$ respectively; these links are not integrated over.  Then
\bea
e^{\D S_P} &=&  {\int  DU_k  D\phi ~  e^{S'_L} \over \int  DU_k  D\phi ~  e^{S''_L} }
\non \\ 
&=& {\int  DU_k  D\phi ~  \exp[S'_L-S''_L] e^{S''_L} \over \int  DU_k  D\phi ~  e^{S''_L} }
\non \\
&=& \Bigl\langle  \exp[S'_L-S''_L] \Bigr\rangle''
\eea
where $\langle ... \rangle''$ indicates that the VEV is to be taken in the probability measure
\bea
{e^{S''_L} \over  \int  DU_k  D\phi ~  e^{S''_L} }
\eea

    As mentioned in the previous section, for the SU(2) gauge group 
the effective action $S_P$ depends only on the trace ${P_{\vx} = \oh \tr[U_{\vx}]}$ of Polyakov line holonomies, so we may consider two configurations in which the Fourier decomposition \rf{dft} of Polyakov lines in each configuration
differ only by $\D a_\vk$ in the amplitude of a particular Fourier component.  In that case we can estimate the derivative
by a finite difference
\bea
\left( {\pa S_P \over \pa a_{\vk}} \right)_{a_{\vk}=\a} \approx   {\D S_P \over \D a_{\vk}}           
\label{deriv}
\eea
The remaining Fourier components, which are the same for Polyakov lines in $U'$ and $U''$, are derived from a thermalized lattice configuration.  The procedure is to generate a thermalized configuration $U_\m(\vx,t)$ by the usual lattice Monte Carlo method, calculate the associated Polyakov line holonomies
\bea
U_\vx \equiv U_0(\vx,1) U_0(\vx,2)...U_0(\vx,N_t)
\eea
and carry out the Fourier decomposition of the corresponding Polyakov lines \rf{dft}.   Then pick a particular wavenumber $\vk$
and set $a_\vk=0$.  Denote the modified Polyakov lines as $\tP_\vx$.  We then construct two other Polyakov line configurations   
\bea
P'_\vx &=& \a \cos(\vk \cdot \vx) + f \tP_\vx
\non \\
P''_\vx &=& (\a + \D a_\vk)\cos(\vk \cdot \vx) + f \tP_\vx
\eea
along with corresponding holonomies $U'_\vx, U''_\vx$ which give rise to these Polyakov lines.  The constant 
$f \approx 1-\a$,  is chosen to ensure that the absolute values of $P'_\vx$ and $P''_\vx$ are $\le 1$.   We then compute \rf{deriv} by the relative weights approach. 

    For a detailed exposition of the relative weights method, including noise reduction and the precise way in which we choose $U'_\vx, U''_\vx$ and $f$, the reader is referred to ref.\ \cite{Greensite:2013yd}. 
    
\section{\label{sec:results}Results}

   We have carried out simulations of pure SU(2) lattice gauge theory on a $16^3 \times 4$ lattice volume at the following
$\b$ values:
\bea
\b = 1.2, 1.4, 1.6, 1.8, 2.0, 2.1, 2.2, 2.25, 2.3
\eea
All of these values lie inside the confined phase with the exception of $\b=2.3$, which lies essentially right at the deconfinement 
transition for $N_t=4$ (the precise transition point is $\b_c=2.2986(6)$ \cite{Fingberg:1992ju}).  We also calculate the derivative \rf{deriv} at $\a$ values
\bea
\a = 0.05, 0.10, 0.15, 0.20
\eea
and for a range of lattice momenta $0 \le k_L \le 2.9$, where
\bea
         k_L = \sqrt{4 \sum_{i=1}^3 \sin^2(\oh k_i)}
\eea
The components $k_i$ of the wavevector $\vk$ are $k_i = {2\pi \over L} m_i$, with $L=16$ the spatial extension of the lattice,
and $m_i=$ integers (including zero).

\begin{figure}[t!]
\subfigure[~$\b=1.2$]  
{   
 \label{scaling120}
 \includegraphics[scale=0.6]{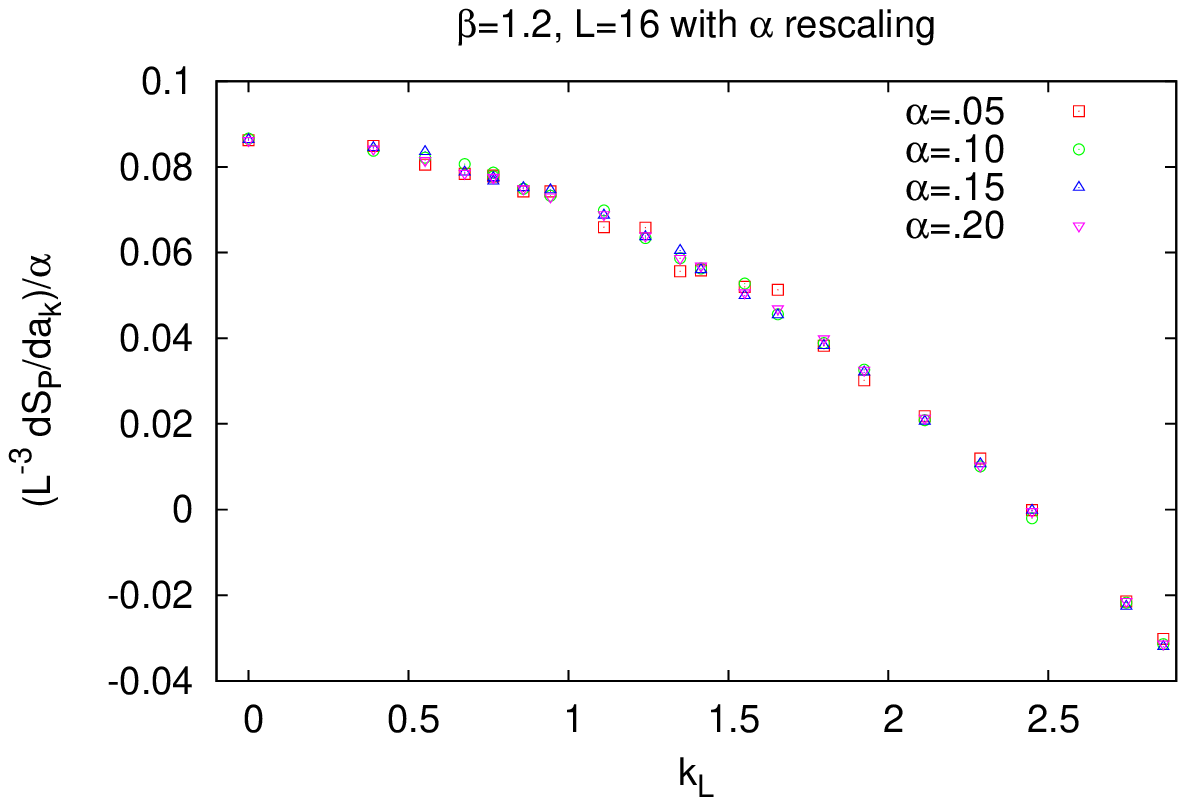}
}
\subfigure[~$\b=1.4$]  
{   
 \label{scaling140}
 \includegraphics[scale=0.6]{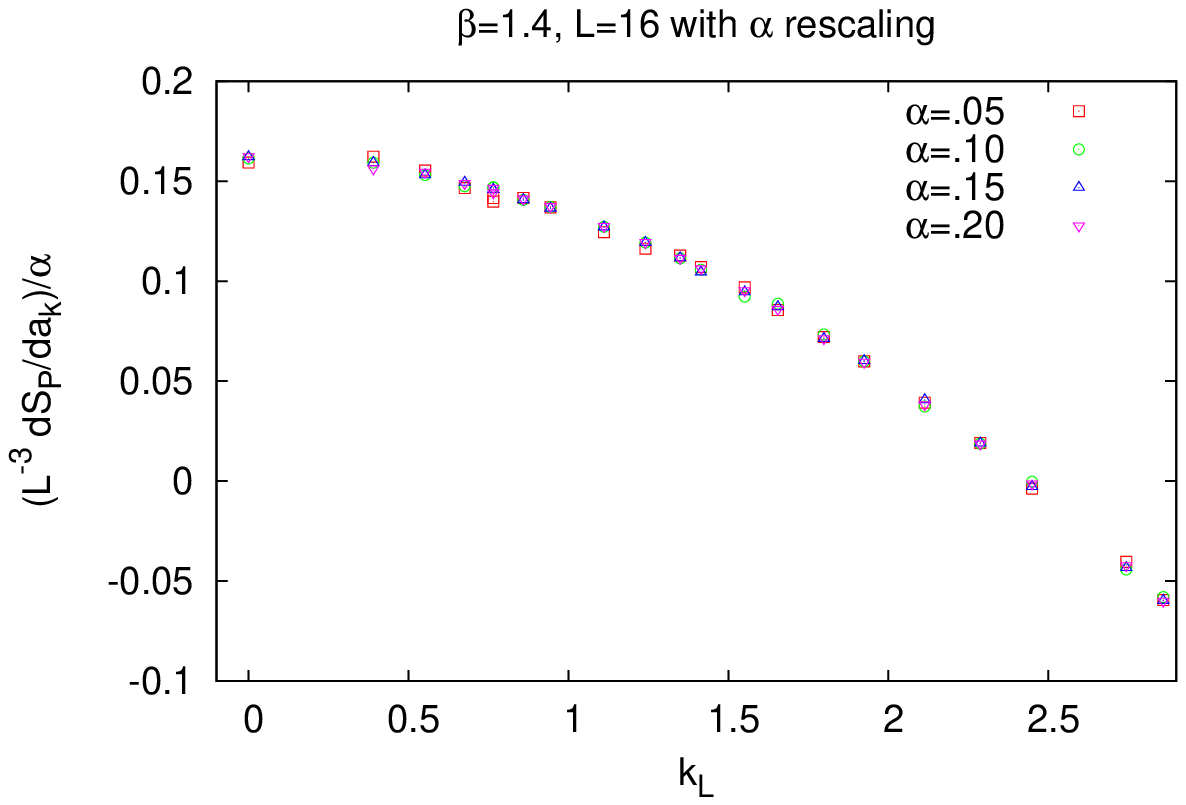}
}
\subfigure[~$\b=1.6$]  
{   
 \label{scaling160}
 \includegraphics[scale=0.6]{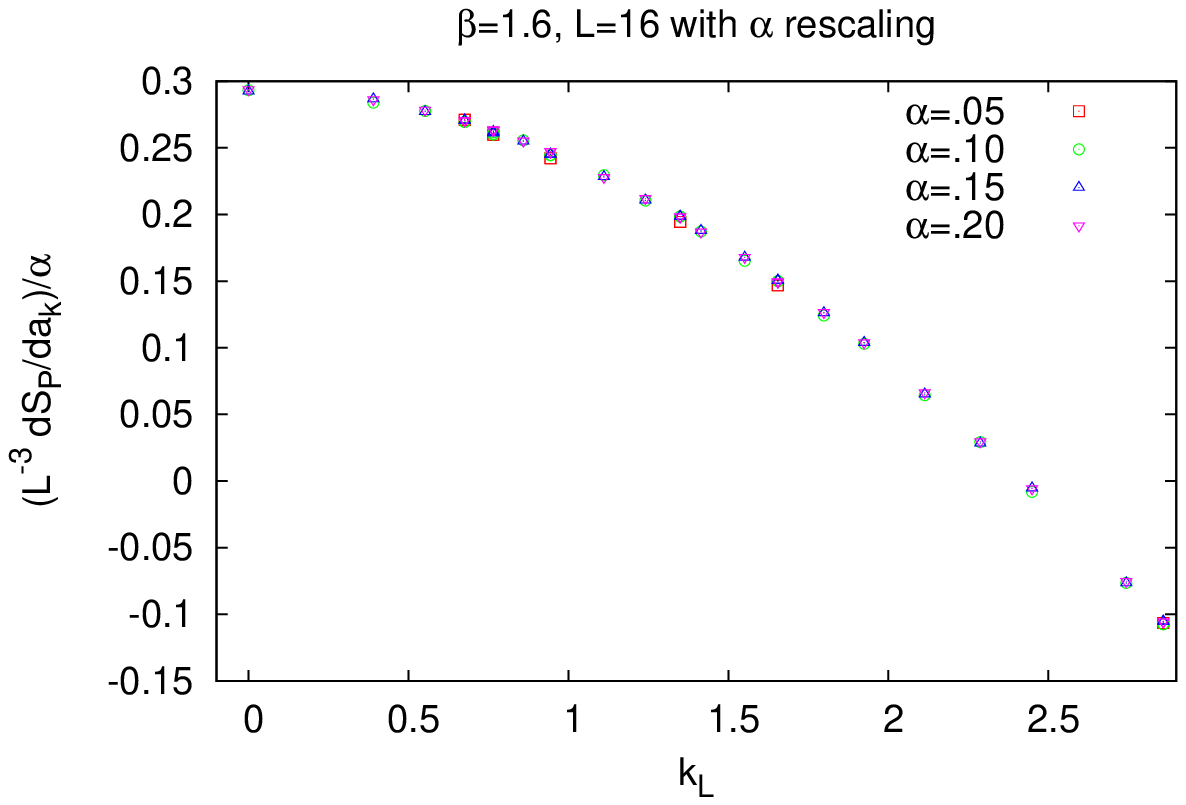}
}
\subfigure[~$\b=1.8$]  
{   
 \label{scaling180}
 \includegraphics[scale=0.6]{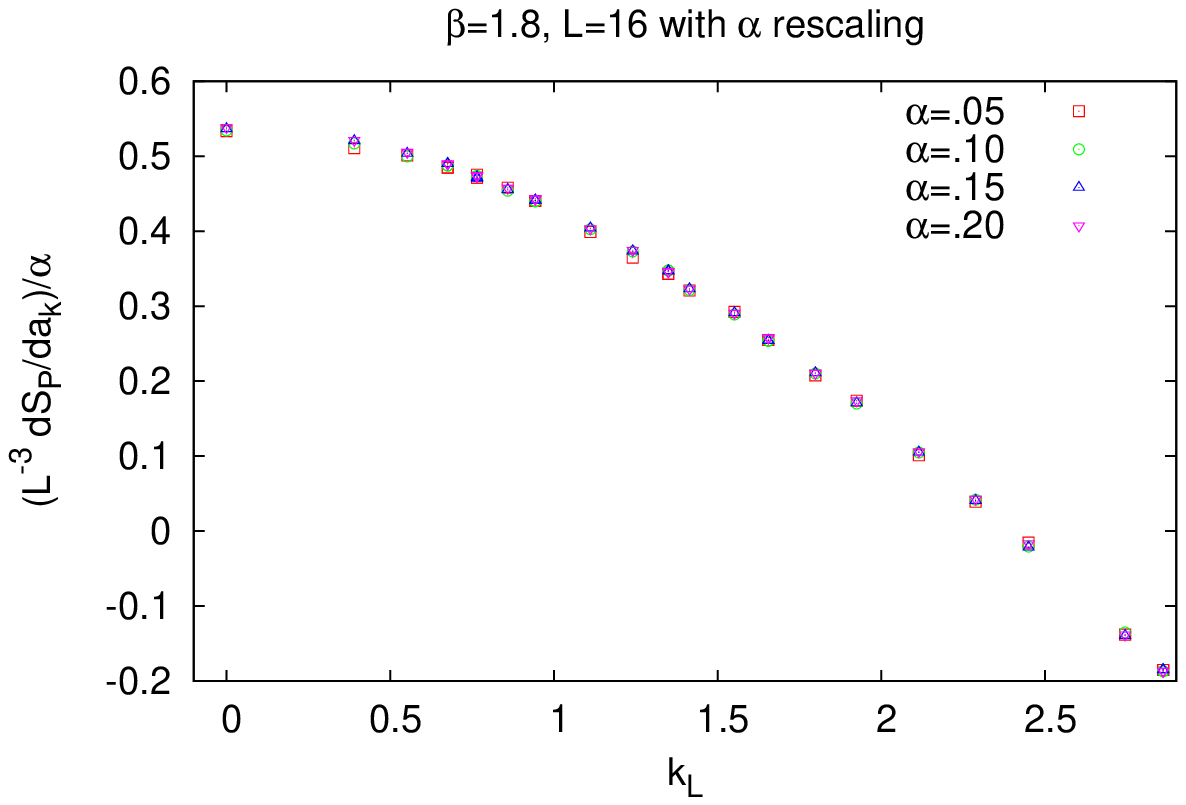}
}
\caption{Derivatives of $S_P$ with respect to Fourier components $a_\vk$, evaluated at $a_\vk=\a$, and divided by $\a$,
for intermediate/strong-coupling values of $\b=1.2-1.8$.  The data, obtained by the relative weights method, is plotted vs.\ $k_L$.  The point displayed at $k_L=0$ is the data value divided by two, for reasons explained in the text.}
\label{scaling}
\end{figure}

\begin{figure}[h!]
\subfigure[~$\b=2.0$]  
{   
 \label{scaling200}
 \includegraphics[scale=0.6]{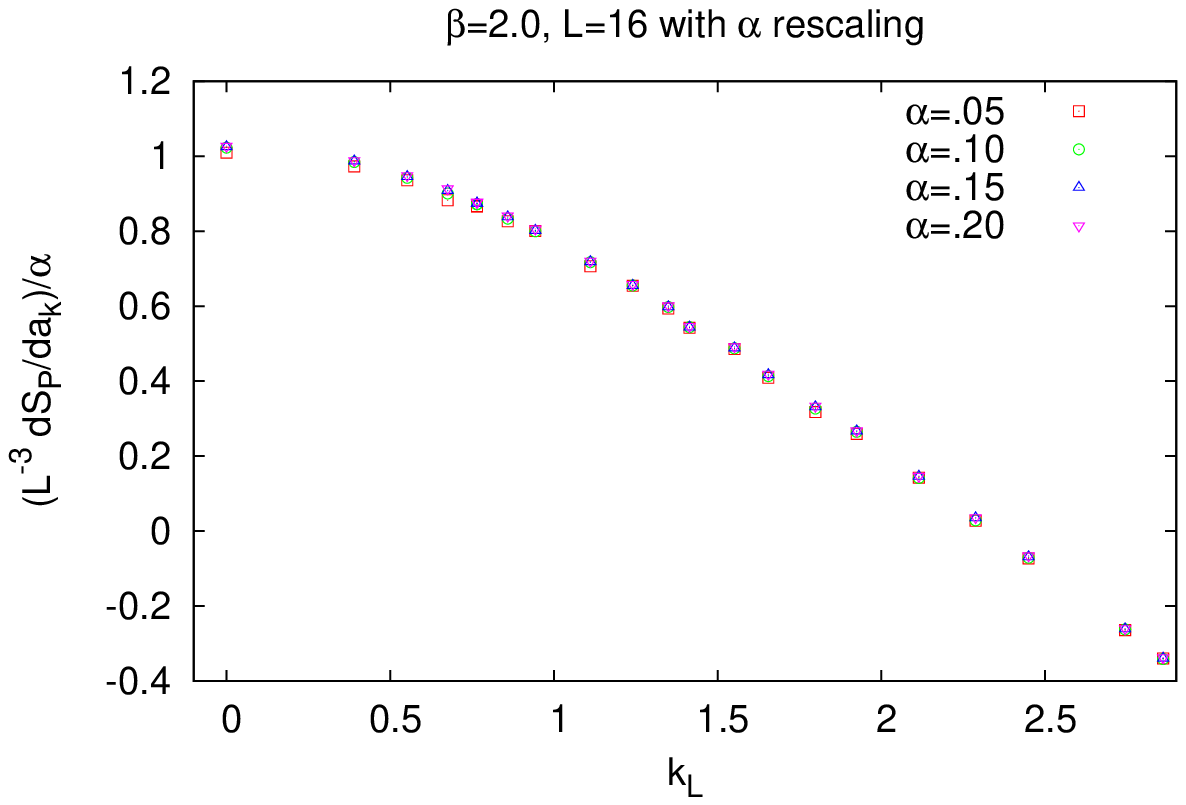}
}
\subfigure[~$\b=2.1$]  
{   
 \label{scaling210}
 \includegraphics[scale=0.6]{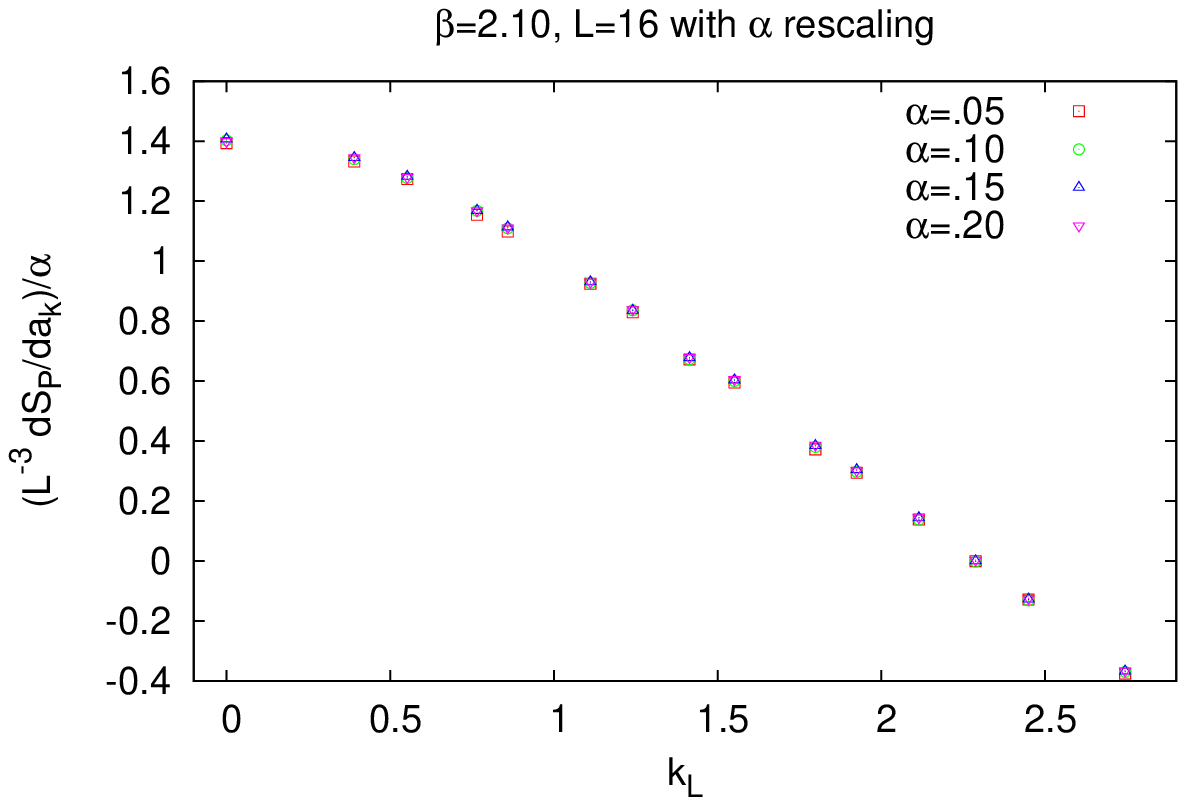}
}
\subfigure[~$\b=2.2$]  
{   
 \label{scaling220}
 \includegraphics[scale=0.6]{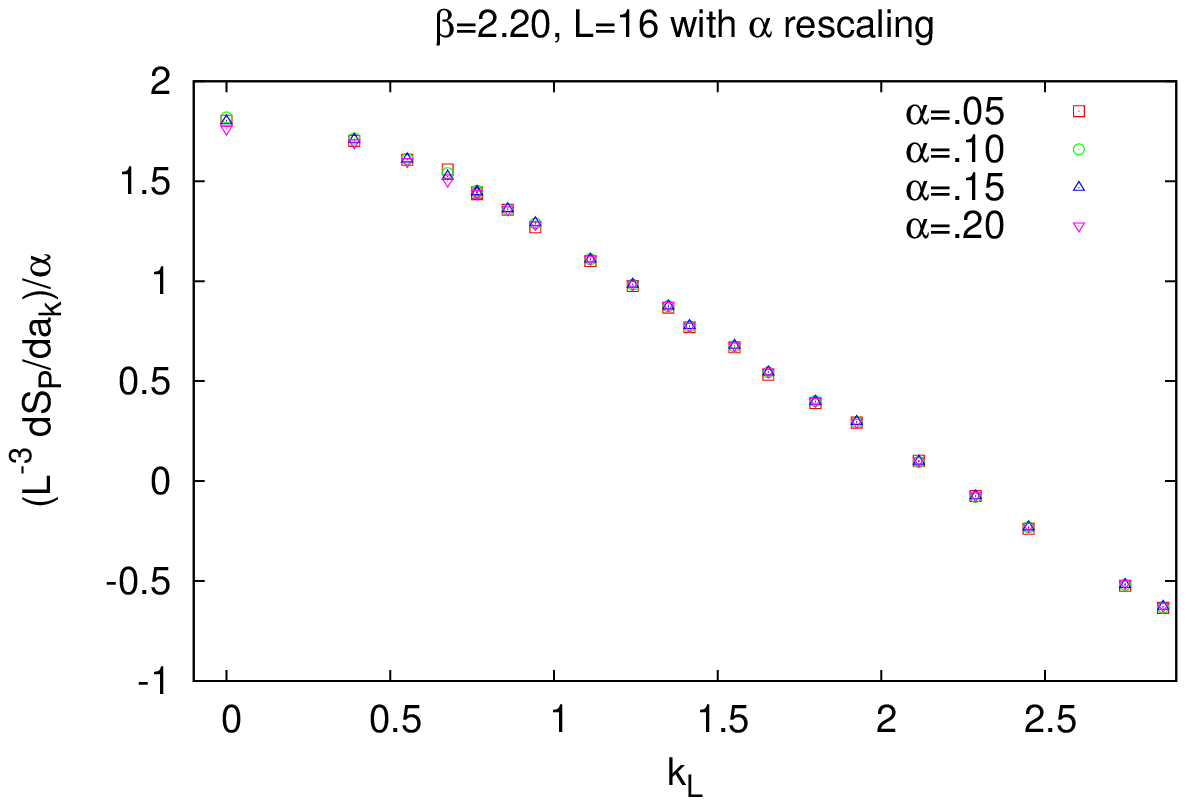}
}
\subfigure[~$\b=2.25$]  
{   
 \label{scaling225}
 \includegraphics[scale=0.6]{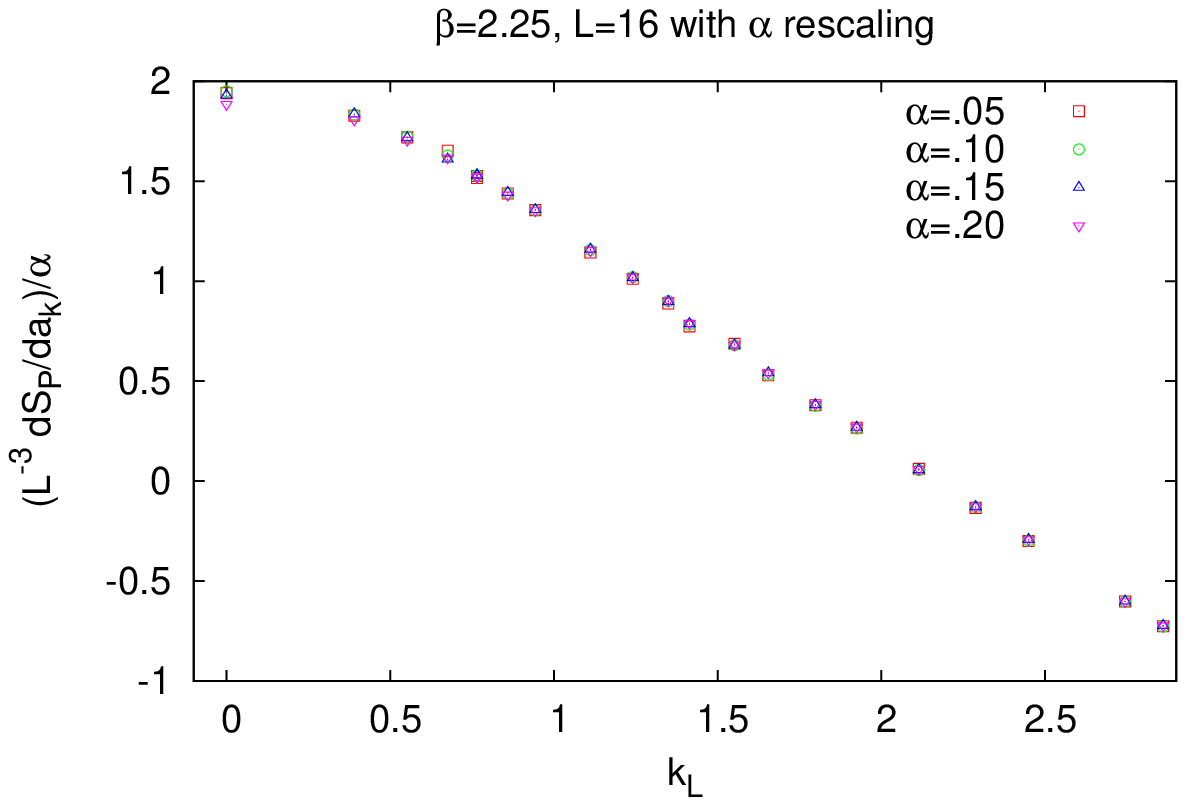}
}
\subfigure[~$\b=2.3$]  
{   
 \label{scaling230}
 \includegraphics[scale=0.6]{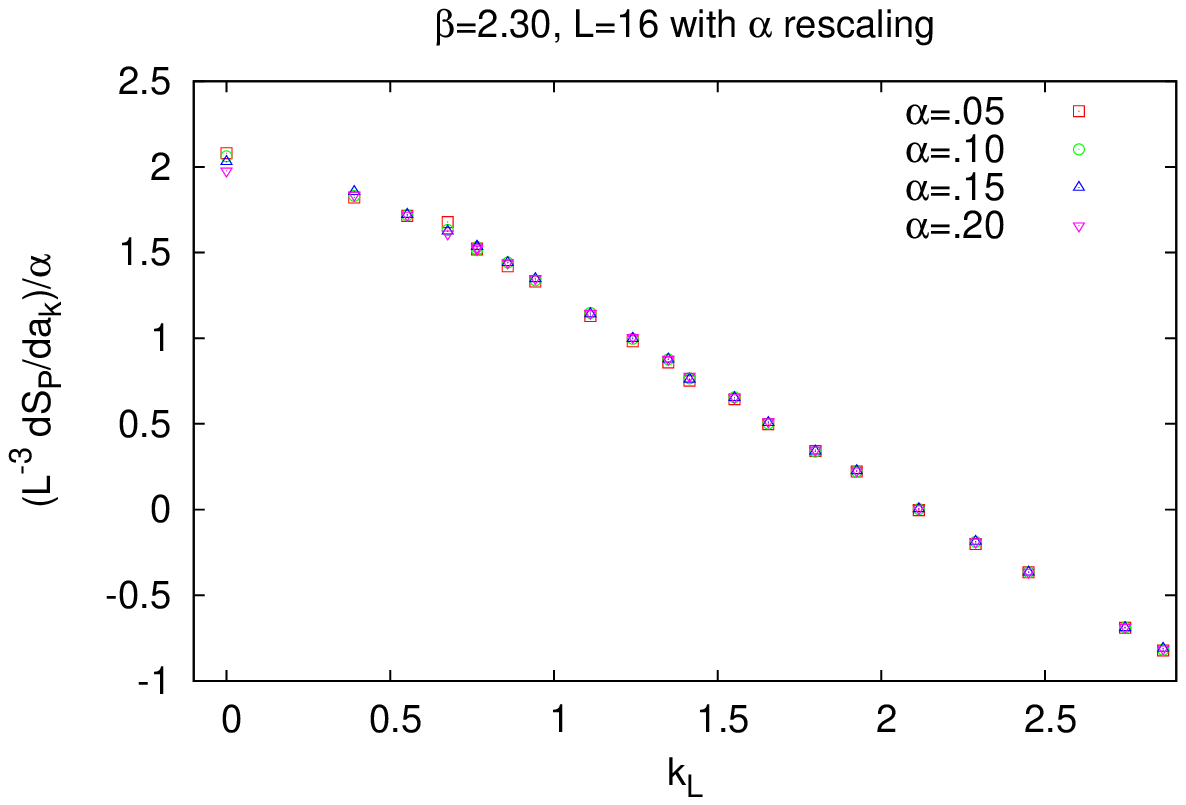}
}
\caption{Same as Fig.\ \ref{scaling}, for intermediate/weak-coupling values $\b=2.0-2.3$.}  
\label{scaling1}
\end{figure}

    The striking fact is that, in all cases, the derivative \rf{deriv} of $S_P$ is linear in $\a$.  This is evident when we plot
\bea
         {1\over \a}  {1\over L^3} \left( {\pa S_P \over \pa a_{\vk}} \right)_{a_{\vk}=\a} ~~~\mbox{vs.}~~~ k_L
\eea
as shown in Fig.\ \ref{scaling} ($\b=1.2 - 1.8$) and Fig.\ \ref{scaling1} ($\b=2.0 - 2.3$).  
In these figures the data points displayed at $k_L=0$ are actually equal to the data values divided by a factor of two, for reasons we will explain.  The important feature to notice in each of these plots
is that at any given $\b$ and $k_L$, the data points at different $\a$ essentially coincide.  This means that the first derivative of
$S_P$ is linear in $\a$, and it follows that $S_P$ itself is quadratic in $a_\vk$ for any $\vk$.  As a consequence, $S_P$ will also be quadratic in the position space variables (i.e.\ the Polyakov lines $P_\vx$) and we can write this effective action in the form 
\rf{SP1}.  The problem is then to determine $c_1,c_2$ and $Q(\vx-\vy)$ from the data.   

Let $\tQ(k_L)$ be the finite Fourier transform of $Q(\vx)$.  This leads to derivatives
\bea
{1\over L^3} \left({d S_P[U_\vx(a_{\vk})] \over da_{\vk}}\right)_{a_{\vk}=\a} &=& \left\{ 
   \begin{array}{cc}
      \a(\oh c_1 - 2c_2 \tQ(k_L)) & k_L \ne 0 \cr
       & \cr
      2\a(\oh c_1 - 2c_2 \tQ(0))   &   k_L=0 \end{array} \right. \ .
\label{dS_P}
\eea       
The relative factor of two in the $k_L=0$ and $k_L > 0$ cases is due to the fact that $\sum_{\vx} 1 = L^3$, while
$\sum_{\vx} \cos^2(\vk \cdot \vx) = \oh L^3$.  The $k_L > 0$ data should extrapolate, as $k_L \ra 0$, to a value which
is half the result at $k_L=0$.  For this reason, in Fig.\ \ref{scaling} and subsequent figures, the point shown at $k_L=0$  is the data value at $k_L=0$ divided by two.             

   Let us begin with the data for lattice couplings $\b$ in the intermediate/weak-coupling regime $2.0 \le \b \le 2.3$ shown
in Fig.\ \ref{scaling1}.  In addition to the $\a$-independence, what is striking about this data is that it is clearly linear for most of the range of lattice momenta.  If the data were linear for the entire range, then we would have
\bea
           S_P =  \oh c_1 \sum_{\vx} P^2_{\vx} -  2c_2 \sum_{\vx \vy} P_{\vx} \Bigl(\sqrt{-\nabla_L^2}\Bigr)_{\vx \vy} P_{\vy} \ ,
\label{SPlin}
\eea
In this case the kernel is $Q(\vx-\vy) = (\sqrt{-\nabla_L^2})_{\vx \vy}$ which translates in momentum space to $\tQ(\vk)=k_L$.  This corresponds to the linear behavior seen in Fig.\ \ref{scaling1}.  
However, the action \rf{SPlin} has a long-range coupling between Polyakov lines which, in the first place, is inconsistent with the non-linear behavior seen at low $k_L$, and in the second place would violate one of the assumptions of the Svetitsky-Yaffe conjecture \cite{Svetitsky:1985ye}, which postulates only finite-range couplings in the effective action.  A simple finite range ansatz for $Q(\vx-\vy)$, having the linear behavior in momentum space seen at high $k_L$, is the expression \rf{Q} proposed in ref.\ \cite{Greensite:2013yd}.   Since $\tQ(\vk) \ra k_L$ at high $k_L$, the constants $c_1,c_2$ are determined from a linear fit of the data in the linear (higher-momentum) regime to the expression~\footnote{The precise choice of interval in $k_L$ to carry out the linear fit is a potential source of systematic error in $c_1,c_2$.  In practice we choose a low-momentum cutoff which minimizes the $\chi^2$ value of the linear fit.} 
\bea
{1\over \a} {1\over L^3} \left({d S_P[U_\vx(a_{\vk})] \over da_{\vk}}\right)_{a_{\vk}=\a} = \oh c_1 - 2c_2 k_L
\eea
The action \rf{SP1} is then well-defined, given the finite range cutoff $r_{max}$.  

     To determine $r_{max}$, we look for the value which satisfies the lower ($k_L=0$) identity in \rf{dS_P}, i.e.
\bea
 {1\over \a} {1\over L^3} \left({d S_P \over da_0}\right)_{a_0=\a} =   c_1 - 4c_2 \tQ(0)
\label{rmax}
\eea
as accurately as possible.  We cannot satisfy this relationship exactly because, while $\b$ can be varied
continuously, $r_{max}$ cannot, since on the lattice it is the square root of a sum of squared integers.
Nevetheless, if $r_{max}$ is not too
small, we can come close to satisfying the equality.  This is essentially an optimization problem.  For each trial $r_{max}$,
compute $Q(\vx-\vy)$ from \rf{Q}, and then Fourier transform to obtain $\tQ(k_L)$.  The optimal choice of $r_{max}$ is the value
which comes closest to satisfying \rf{rmax}.  We can then compute the right hand side of \rf{dS_P} at all $k_L$, and see how well it fits the data.  The fractional deviation between the left and right hand sides of \rf{rmax}
\bea
   \d = { {1\over \a} {1\over L^3} \left({d S_P \over da_0}\right)_{a_0=\a} -  c_1 - 4c_2 \tQ(0) \over {1\over \a} {1\over L^3} 
   \left({d S_P \over da_0}\right)_{a_0=\a} }
\label{fracdev}
\eea
is on the order of 1\% or less, for $\b > 2.0$, and about 2\% at $\b=2.0$.

\begin{figure}[t!]
\subfigure[~$\b=2.0$]  
{   
 \label{comfit200}
 \includegraphics[scale=0.6]{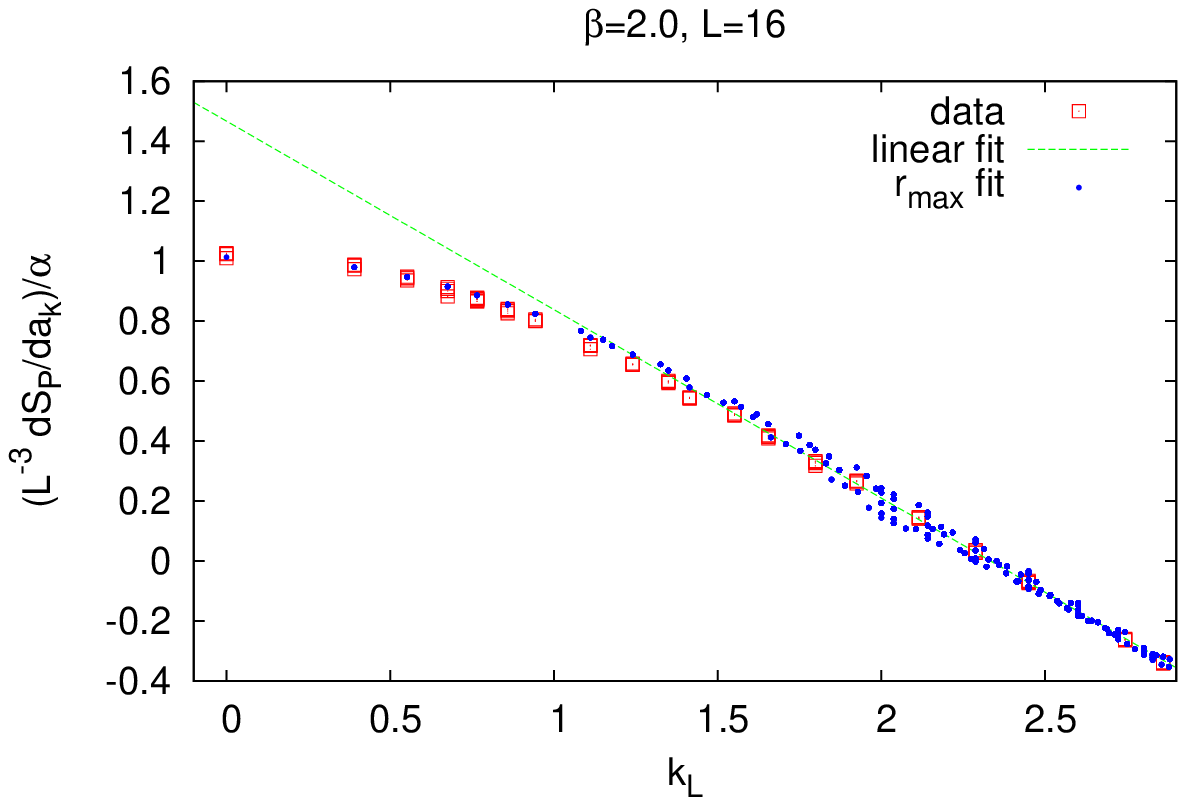}
}
\subfigure[~$\b=2.1$]  
{   
 \label{comfit210}
 \includegraphics[scale=0.6]{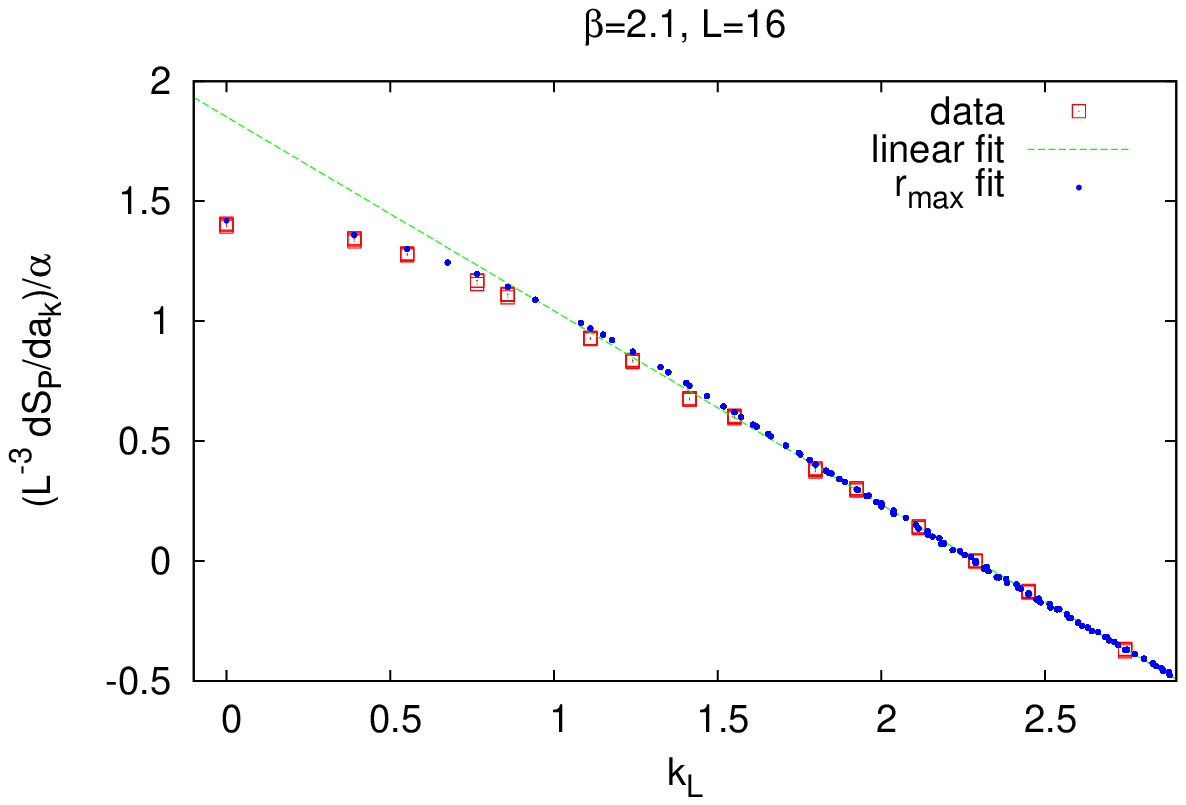}
}
\subfigure[~$\b=2.2$]  
{   
 \label{comfit220}
 \includegraphics[scale=0.6]{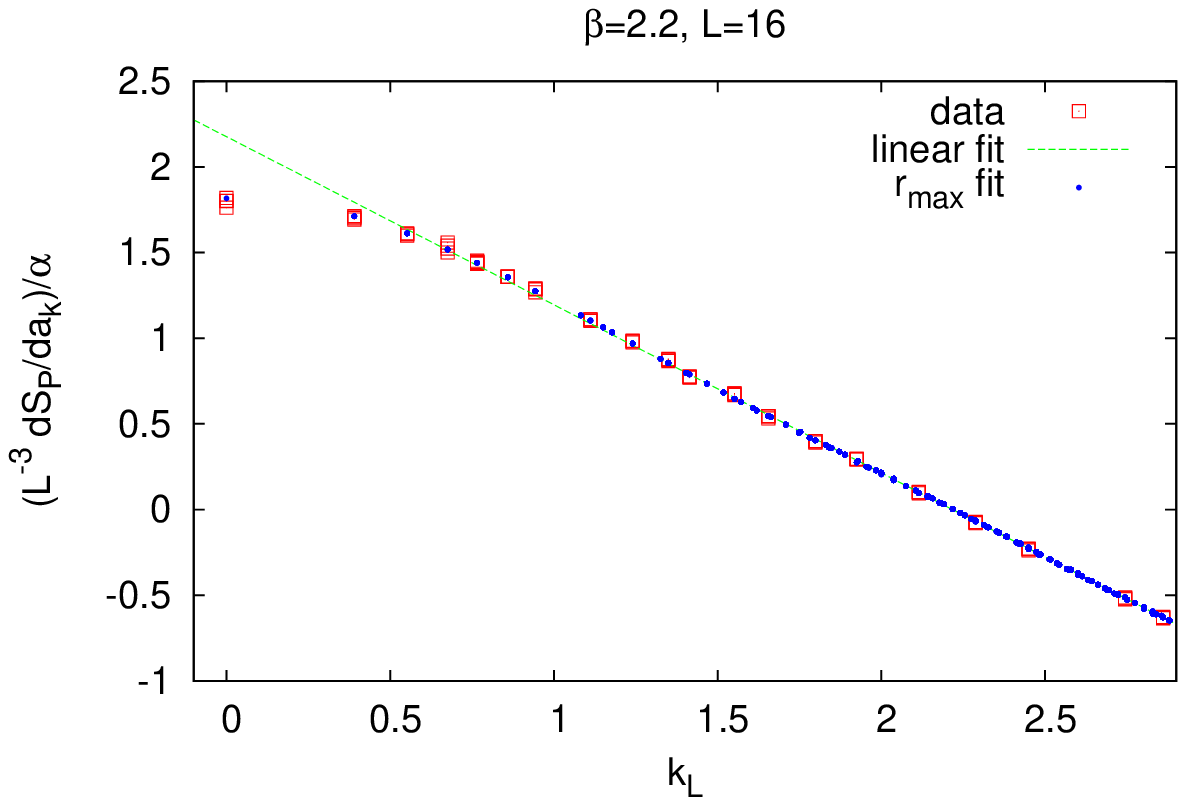}
}
\subfigure[~$\b=2.25$]  
{   
 \label{comfit225}
 \includegraphics[scale=0.6]{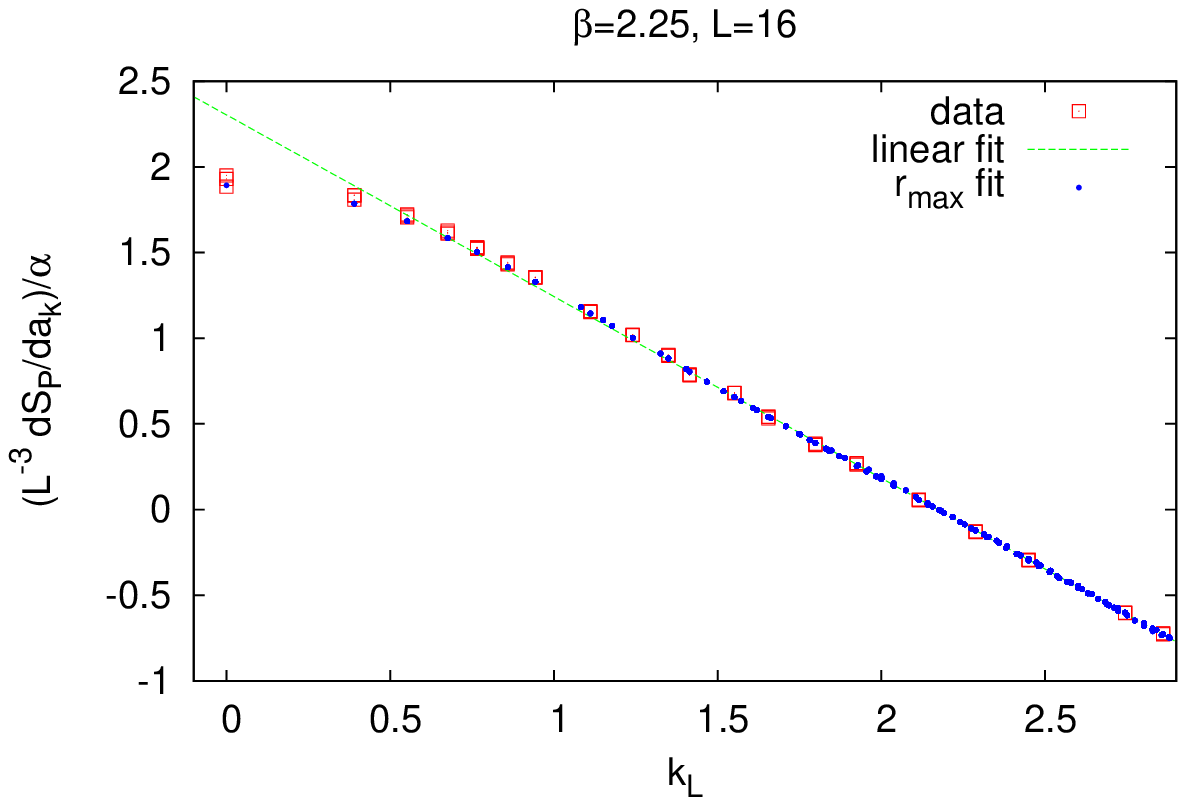}
}
\subfigure[~$\b=2.3$]  
{   
 \label{comfit230}
 \includegraphics[scale=0.6]{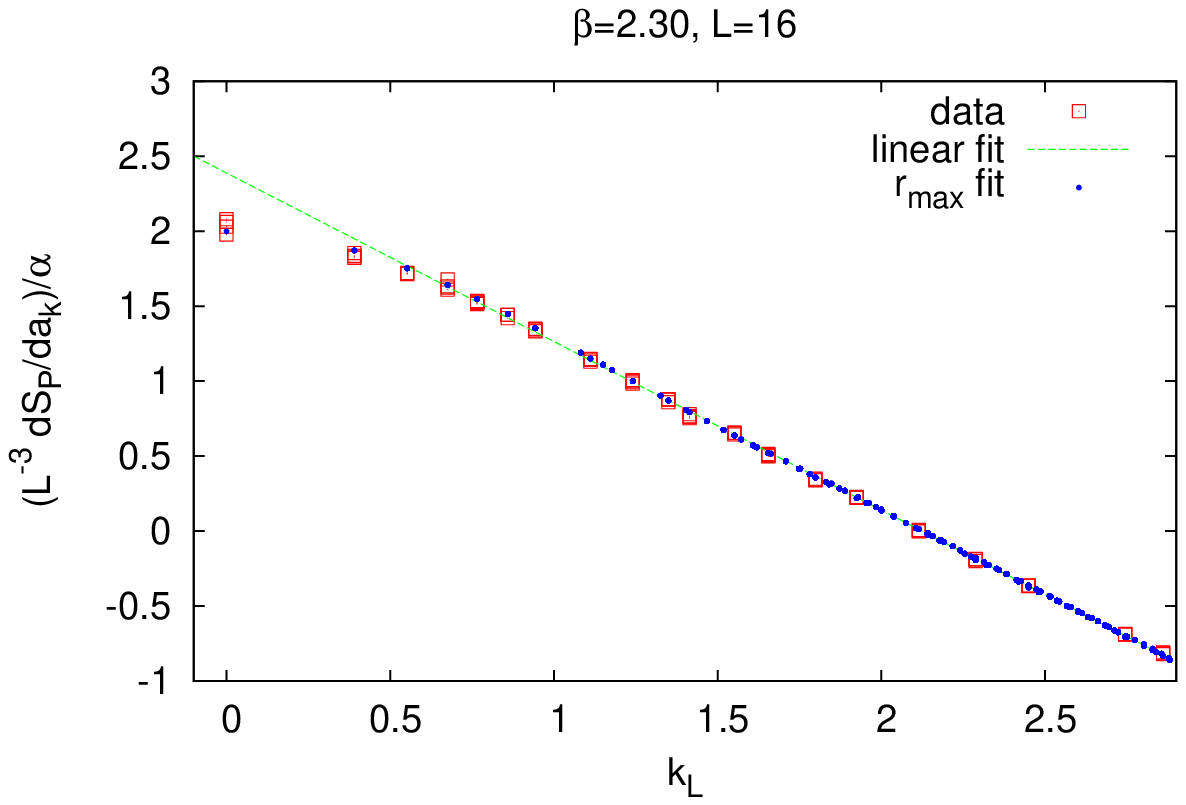}
}
\caption{Comparison of $c_1/2 - 2c_2 \tQ(\vk)$ to the relative weights data, where $\tQ(\vk)$ is the Fourier transform of the kernel \rf{Q}, for couplings $\b=2.0-2.3$. Red squares are data points, blue dots are the $c_1/2 - 2c_2 \tQ(\vk)$ values, and the green line is the linear fit used to determine $c_1,c_2$.}
\label{comfit1}
\end{figure}

    The results, shown in Fig.\ \ref{comfit1}, seem to be quite satisfactory.  The green line is the linear fit used to compute
$c_1, c_2$, the blue dots correspond to $\oh c_1 - 2c_2 \tQ(k)$. The red open squares are the data points already seen in Fig.\ \ref{scaling1}, this time with no distinction on $\a$ values.

\begin{figure}[t!]
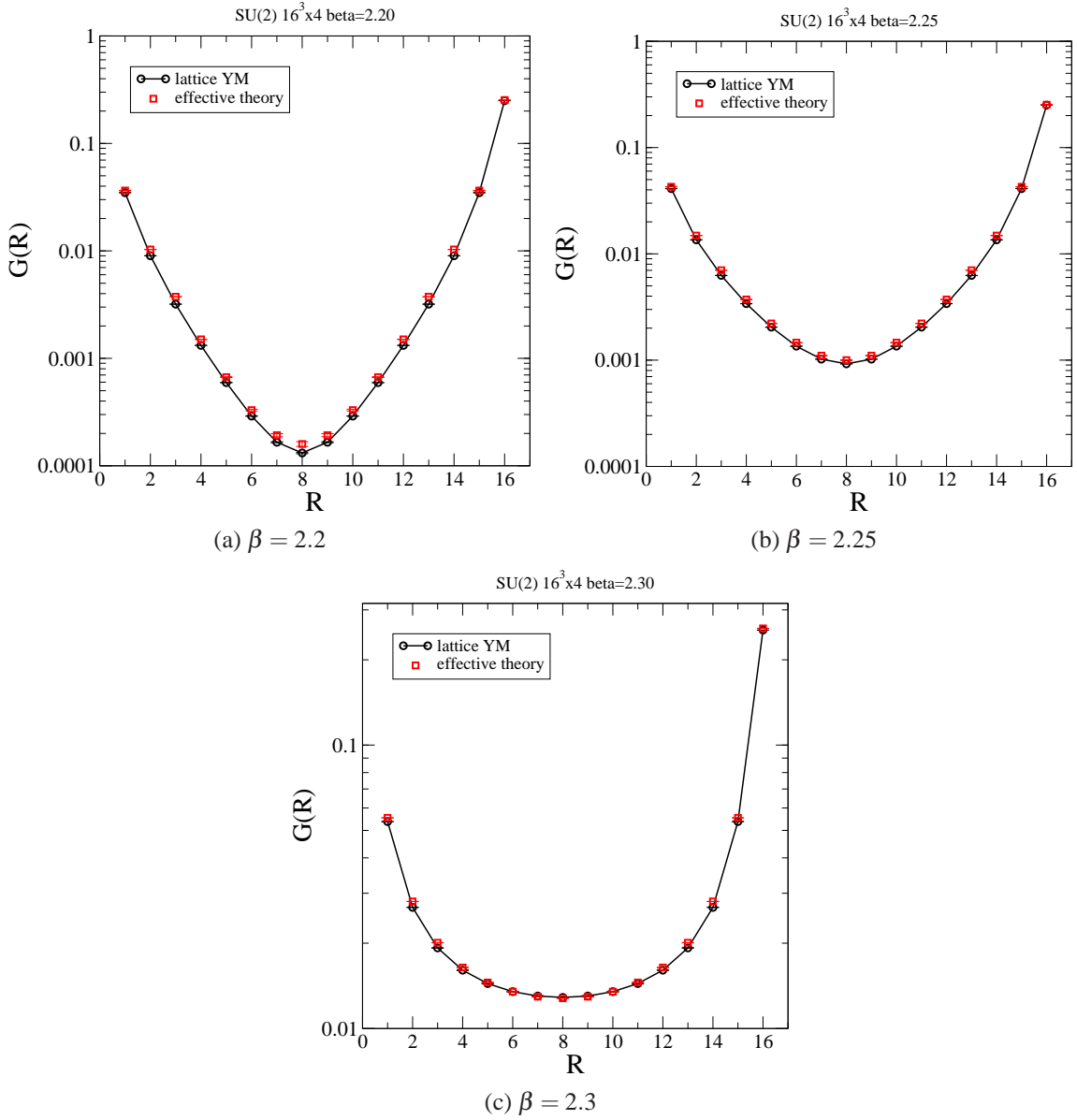

\subfigure[~$\b=2.2$]  
{   
 \label{pot220}
 \includegraphics[scale=0.55]{pot16_220.eps}
}
\subfigure[~$\b=2.25$]  
{   
 \label{pot225}
 \includegraphics[scale=0.4]{pot16_225.eps}
}
\subfigure[~$\b=2.3$]  
{   
 \label{pot230}
 \includegraphics[scale=0.4]{pot16_230.eps}
} 
\caption{Comparison of Polyakov line correlators $G(R)$ computed by simulation of the effective Polyakov line
action, and by simulation of the underlying lattice gauge theory.}  
\label{pot1}
\end{figure}

    The essential test of the proposed effective action \rf{SP1} with kernel \rf{Q} is the comparison in the Polyakov line correlators
\bea
        G(\vx-\vy) = \langle P_\vx P_\vy \rangle
\eea
computed in both the effective theory, and in the underlying lattice gauge theory.  Results for on-axis separations at 
$\b=2.2,2.25,2.3$ are shown in Fig.\ \ref{pot1}.  The correlators in the lattice gauge theory have been calculated using
L{\"u}scher-Weisz noise reduction \cite{Luscher:2001up}.  

\begin{figure}[t!]
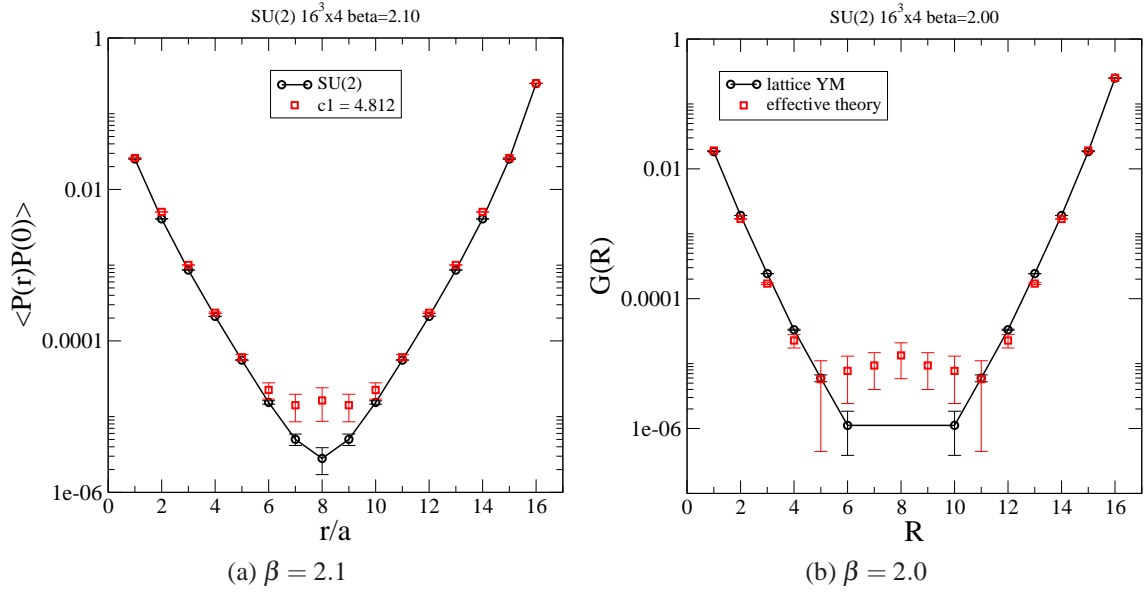

\subfigure[~$\b=2.1$]  
{   
 \label{pot210}
 \includegraphics[scale=0.4]{pot16_210.eps}
} 
\subfigure[~$\b=2.0$]  
{   
 \label{pot200}
 \includegraphics[scale=0.4]{pot16_200.eps}
}
\caption{Same as Fig.\ \ref{pot1}, for $\b=2.0,2.1$.  Note the plateau, in the correlator of the effective theory, around
\ $G(R)=10^{-5}$.}
\label{pot2}
\end{figure}

    The agreement between the correlators in the effective theory and the lattice gauge theory is remarkably accurate,
down to values of $G(\vx-\vy) \sim 10^{-5}$.  However, as $\b$ is reduced, we find that the data in the effective theory
seems to flatten out at around $10^{-5}$, as can be seen in Fig.\ \ref{pot2}. This flattening may simply be a finite volume
effect in the numerical simulation of the effective theory.  The argument is as follows:  For separations $R=|\vx-\vy|$ which are several times greater than a correlation length, say for $R>R_{max}$, Polyakov lines are almost uncorrelated. Let us denote
the average value of a Polyakov line in a given thermalized configuration as
\bea
\overline{P_\vx} = {1\over L^3} \sum_\vx P_x
\eea
and also define
\bea
\overline{(P_\vx P_\vy)}_{R>R_{max}} \equiv { \sum_\vx \sum_\vy P_\vx P_\vy \theta(R-R_{max}) \over
\sum_\vx \sum_\vy \theta(R-R_{max}) }
\eea
where $\theta(x)$ is the Heaviside theta function.
Now for $R>R_{max}$ Polyakov lines $P_\vx$ and $P_\vy$ are essentially uncorrelated,  and if the lattice volume
$L^3$ is much greater than $R_{max}^3$, we may approximate $P_\vx$ and $P_\vy$ in the double sum by their average values in the configuration.  But this means that in each configuration
\bea
\overline{(P_\vx P_\vy)}_{R>R_{max}} \approx \Bigl( \overline{P_\vx} \Bigr)^2
\eea
On any given site, the typical magnitude of $P_x$ is perhaps on the order of 0.2, which means that the average value on the lattice, in a typical configuration, will be on the order of $\overline{P_\vx} \sim \pm 0.2/L^{3/2}$.  For $L=16$, this gives us an estimate of 
\bea
\overline{(P_\vx P_\vy)}_{R>R_{max}} \approx 10^{-5}
\eea
in any thermalized configuration in the effective theory on a $16^3$ lattice volume.  This may explain why the measured values
of $G(\vx-\vy)$ seem to plateau at around $10^{-5}$ at these lattice volumes.

\begin{figure}[t!]
\subfigure[~$\b=1.2$]  
{   
 \label{comfit120}
 \includegraphics[scale=0.6]{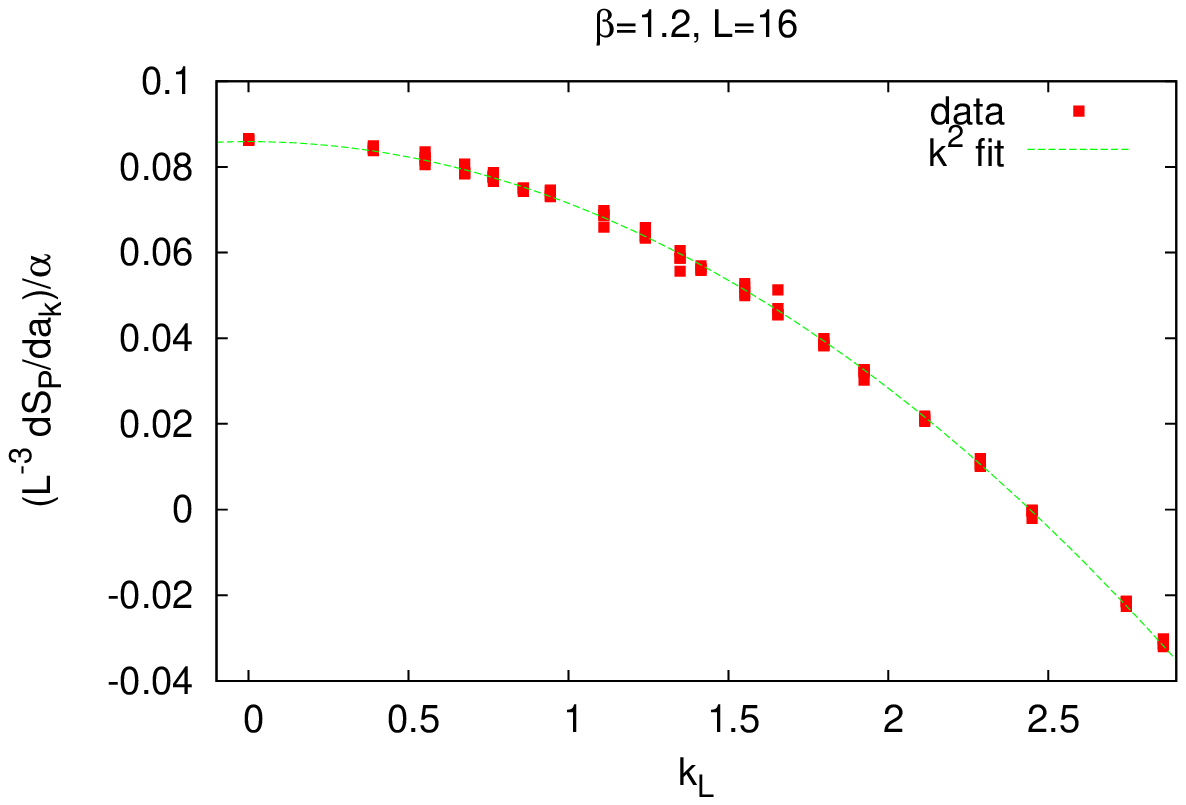}
}
\subfigure[~$\b=1.4$]  
{   
 \label{comfit140}
 \includegraphics[scale=0.6]{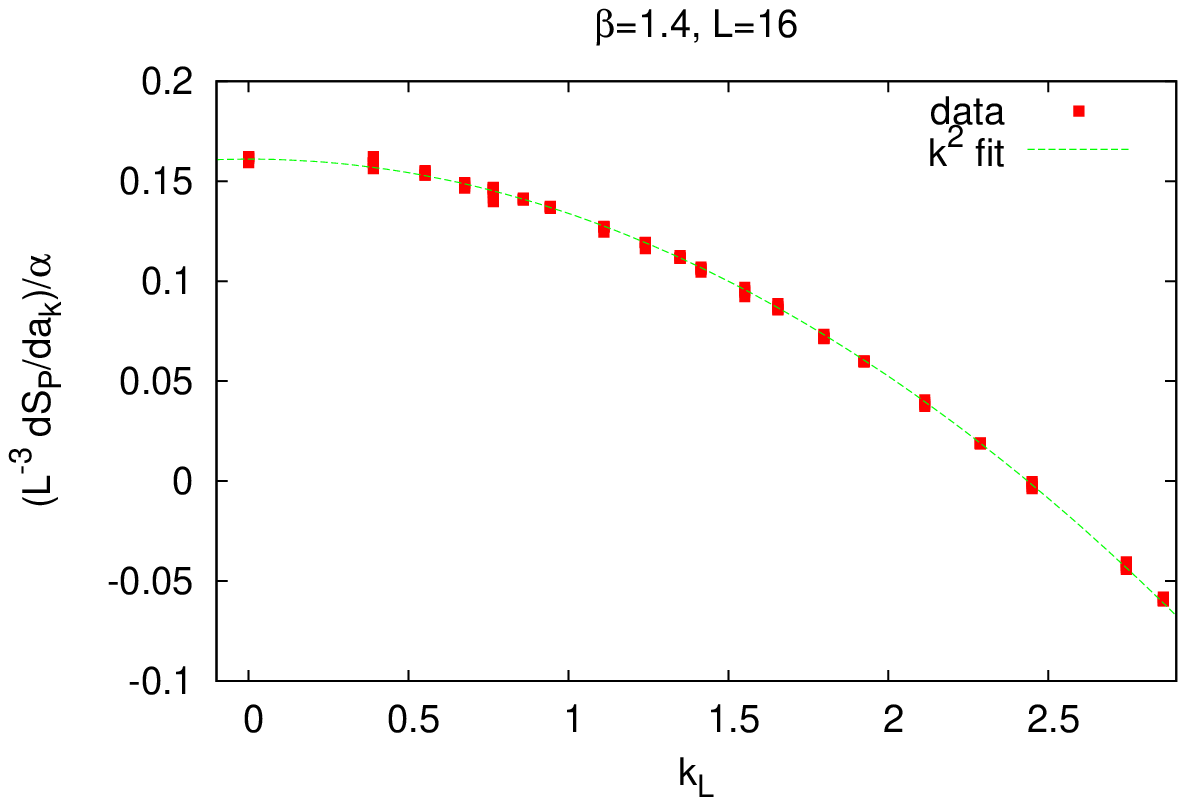}
}
\caption{Comparison of the best fit $c_1/2 - 2c_2 k_L^2$ to the relative weights data at strong-couplings
$\b=1.2,1.4$.} 
\label{comfit2}
\end{figure}

\begin{figure}[h!]
\subfigure[~$\b=1.6$]  
{   
 \label{comfit160}
 \includegraphics[scale=0.6]{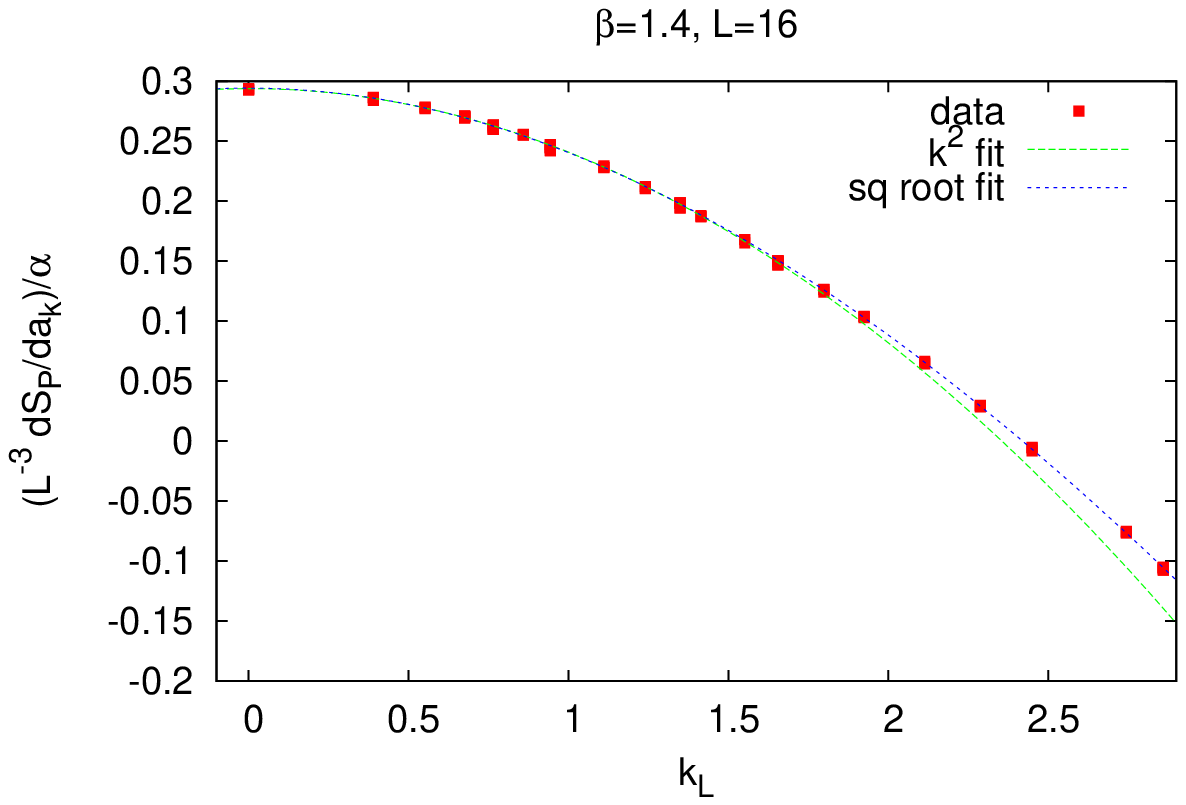}
}
\subfigure[~$\b=1.8$]  
{   
 \label{comfit180}
 \includegraphics[scale=0.6]{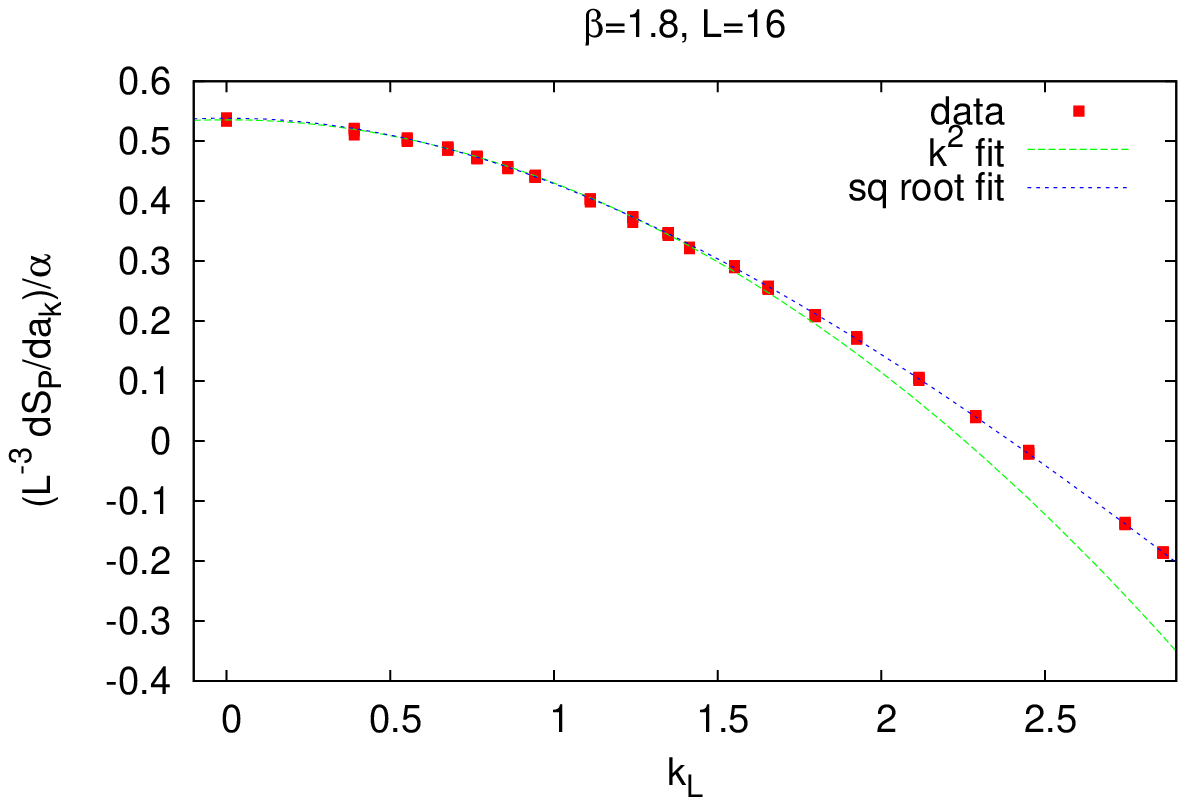}
}
\caption{Comparison of the best fit (blue line)  of the square-root ansatz $c_1/2 - 2c_2 \sqrt{k_L^2+m^2}$ to the relative weights data at strong/intermediate-couplings $\b=1.6,1.8$.  Also shown is a fit to $c_1/2 - 2c_2 k_L^2$ at lower momenta (green line).} 
\label{comfit3}
\end{figure}

    Reducing the lattice coupling below $\b=2.0$, we enter the regime of strong couplings.  Our results for
\bea
         {1\over \a}  {1\over L^3} \left( {\pa S_P \over \pa a_{\vk}} \right)_{a_{\vk}=\a} ~~~\mbox{vs.}~~~ k_L
\eea
at $\b<2.0$ were seen in Fig.\ \ref{scaling}.  In these plots there is much more evidence of curvature, and the part of
the graph which fits a straight line disappears as $\b$ is reduced.  It turns out that the curved section of the plots
are compatible with $\tQ(\vk) = k_L^2$, and in fact this fits the data at $\b=1.2, 1.4$ perfectly, as  we see in Fig.\ \ref{comfit2}.
At $\b=1.6, 1.8$, however, there is still a portion of the data which fits a straight line, and so we need an ansatz for
$\tQ(\vk)$ which interpolates between the quadratic form at small $k_L$ and linear at large $k_L$.  A simple choice is
\bea
\tQ(\vk) = \sqrt{\vk^2 + m^2}
\label{Q1}
\eea
and so we do a best fit of the data at all $k_L$ to the form
\bea
{1\over \a} {1\over L^3} \left({d S_P[U_\vx(a_{\vk})] \over da_{\vk}}\right)_{a_{\vk}=\a} = \oh c_1 - 2c_2  \sqrt{\vk^2 + m^2}
\eea
The resulting fits, for lattice couplings $\b=1.6,1.8$, are shown in Fig.\ \ref{comfit3}.
 
\begin{figure}[h!]
\subfigure[~$\b=1.6$]  
{   
 \label{corr160}
 \includegraphics[scale=0.6]{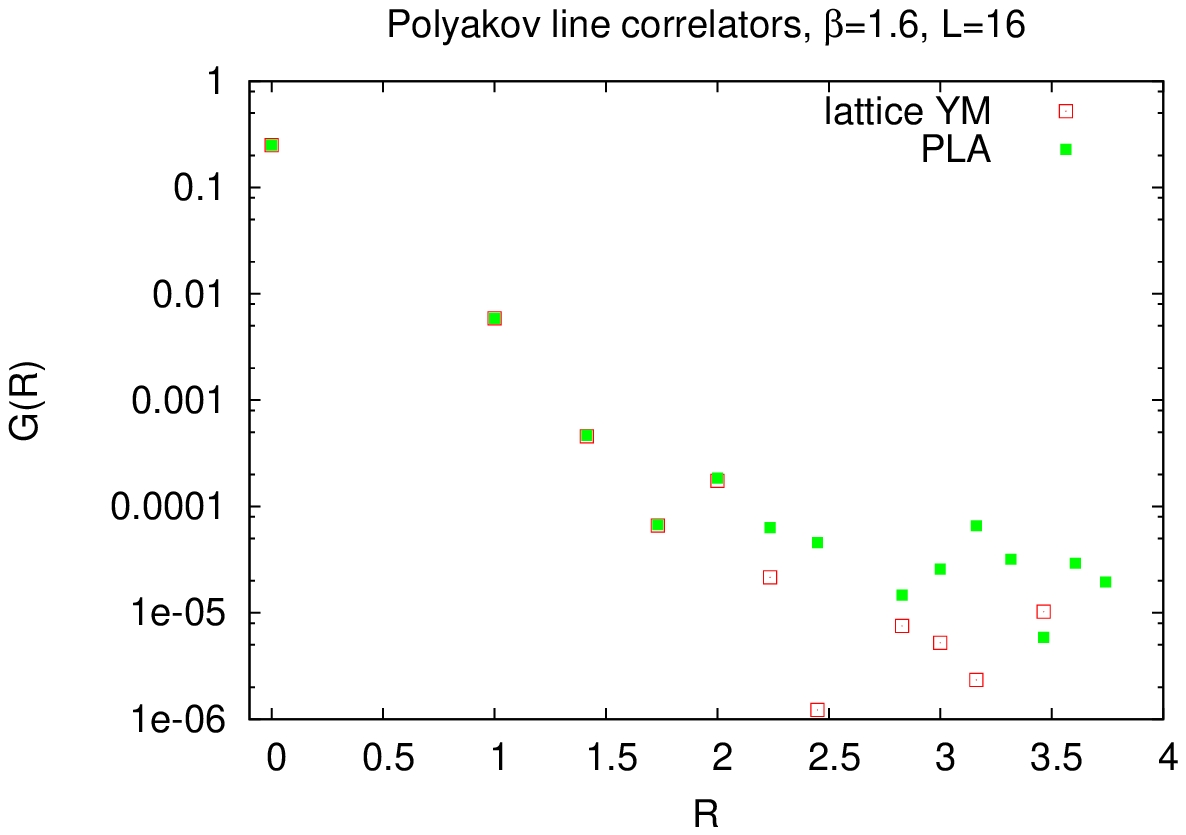}
}
\subfigure[~$\b=1.8$]  
{   
 \label{corr180}
 \includegraphics[scale=0.6]{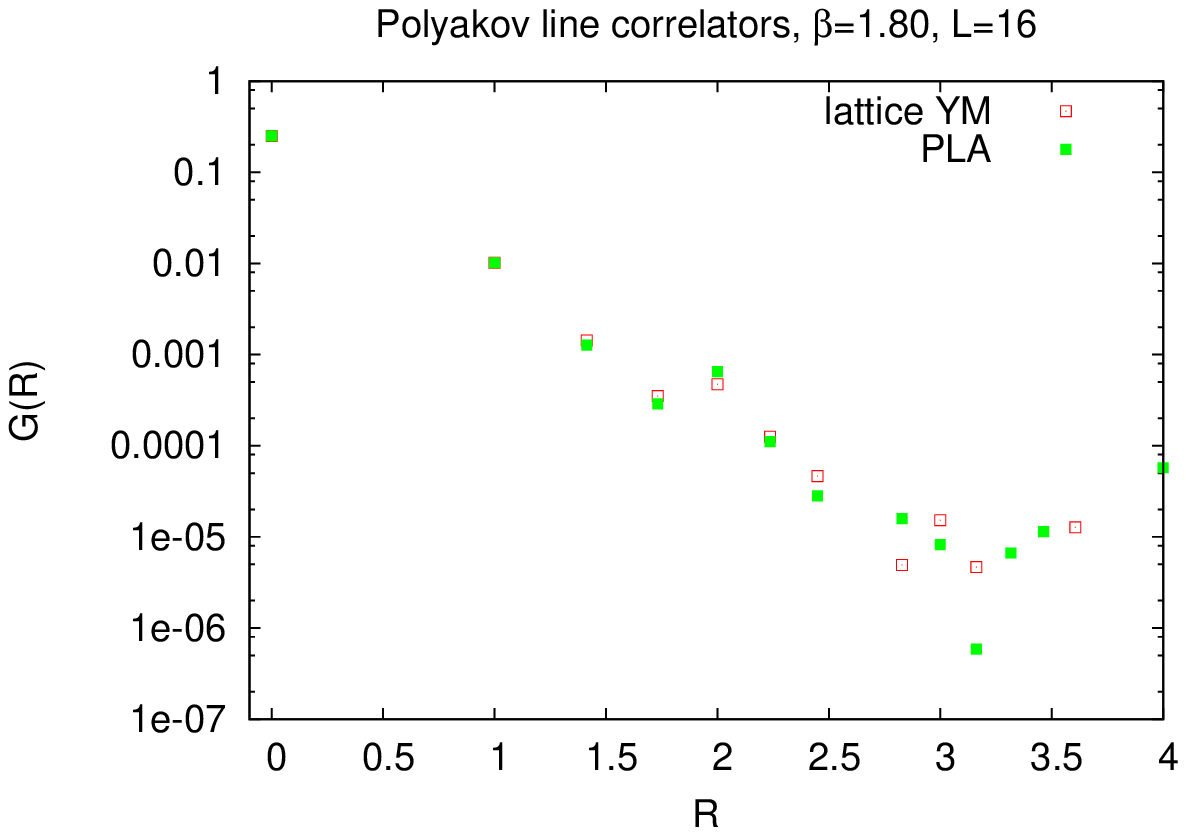}
}
\caption{Comparison of Polyakov line correlators $G(R)$ computed in the effective theory (green dots) and in the underlying lattice gauge theory (red squares), at the intermediate/strong couplings $\b=1.6,1.8$.  These plots include off-axis separations.}
\label{corr1}
\end{figure}

   With the constants $c_1,c_2,m^2$ determined from a best fit to the data, the position space kernel $Q(\vx-\vy)$ is determined
from a fast Fourier transform of $\tQ(\vk)$, and we can again compare the Polyakov line correlators computed in the effective theory with the corresponding correlators computed in the underlying lattice gauge theory.  The results are shown in 
Fig.\ \ref{corr1}, this time including off-axis separations.   Once again, the agreement is very good, although the falloff in position space is so rapid that the $10^{-5}$ plateau sets in at rather small values of $|\vx-\vy|$.  In these figures the Polyakov correlators in the lattice gauge theory are computed in the standard way, without the L{\"u}scher-Weisz noise reduction, and these also seem to show a plateau at the larger lattice separations.

    Since $Q(\vx-\vy)$ corresponding to a square root ansatz does not have a sharp cutoff in the range of $R=|\vx-\vy|$, the corresponding effective action becomes challenging to simulate numerically as $m$ is reduced. We have not yet
investigated this form of the action in the weak coupling regime.

    Finally, at our two strongest couplings, $\b=1.2$ and 1.4, there is no need for an interpolating form, since the 
$\tQ(\vk) = k_L^2$ ansatz fits the data quite well, as seen in Fig.\ \ref{comfit2}.  The data is fit to the form 
\bea
{1\over \a}{1\over L^3} \left({d S_P[U_\vx(a_{\vk})] \over da_{\vk}}\right)_{a_{\vk}=\a} &=& 
      \oh c_1 - 2c_2 k_L^2  
\label{RW}
\eea       
and a short calculation gives us the effective action
\bea
S_P = 4 c_2 \sum_\vx \sum_{i=1}^3   P_\vx P_{\vx+ \ihat} + (\oh c_1 - 12 c_2) \sum_{\vx} P_{\vx}^2
\eea
The best fits to \rf{RW} give the constants
\bea
\begin{array}{lll}
   c_1 = 0.3196(5) &  c_2 = 0.01327(2) & \b=1.4 \cr
   c_2 = 0.1714(2) &  c_2 = 0.007148(8) & \b=1.2 \end{array}
\eea
With these numbers, we find that
\bea
\begin{array}{ll}
   \oh c_1 - 12 c_2 = (-0.56 \pm 0.34) \times 10^{-3}  & ~~~(\b=1.4) \cr
   \oh c_1 - 12 c_2 = (-0.76 \pm 1.4) \times 10^{-4}  & ~~~(\b=1.2) \end{array}
\eea
which are essentially consistent with zero.  Then, comparing the relative weights and strong-coupling results for the PLA
at $\b=1.4$, we find
\bea
S_P = \left\{ \begin{array}{ll}
            0.05308(8)  \sum_\vx \sum_{i=1}^3   P_\vx P_{\vx+ \ihat} & \mbox{relative weights} \cr \cr
            0.0522  \sum_\vx \sum_{i=1}^3   P_\vx P_{\vx+ \ihat} & \mbox{strong coupling} \end{array} \right.  (\b=1.4)
\eea
while at $\b=1.2$,
\bea
S_P = \left\{ \begin{array}{ll}
            0.02859(3)  \sum_\vx \sum_{i=1}^3   P_\vx P_{\vx+ \ihat} & \mbox{relative weights} \cr \cr
            0.02850  \sum_\vx \sum_{i=1}^3   P_\vx P_{\vx+ \ihat} & \mbox{strong coupling} \end{array} \right. (\b=1.2)
\eea
In both cases there appears to be good agreement between the strong-coupling expansion and the relative weights
result.

    To summarize, the effective Polyakov line action is given by \rf{SP1} with the parameters and kernel $Q(\vx-\vy)$
listed in Table \ref{tab1}.  Note that the range of the kernel in lattice units rises from $r_{max} = \sqrt{3} \approx 1.73$, at 
$\b=2.0$, to $r_{max}=\sqrt{13} \approx 3.61$ at the deconfinement transition.

\begin{table}[t!]
\begin{center}
\begin{tabular}{|c|c|c|c|c|c|} \hline
         $ \b $ &  $Q(\vx-\vy)$  & $c_1$& $c_2$ & $m$ & $r_{max}$ \\
\hline
        1.2 & $\Bigl(-\nabla_L^2 \Bigr)_{\vx \vy}$                     & 0.1714(2)  & 0.007148(8)    &    & 1    \\ 
        1.4 & $\Bigl(-\nabla_L^2 \Bigr)_{\vx \vy}$                     & 0.3196(5)  & 0.01327(2)      &    & 1    \\ 
        1.6 & $\Bigl(\sqrt{-\nabla_L^2 + m^2}\Bigr)_{\vx \vy}$ & 4.10(7)   & 0.219(2)   &  4.01(4)  &     \\ 
        1.8 & $\Bigl(\sqrt{-\nabla_L^2 + m^2}\Bigr)_{\vx \vy}$ & 1.969(8) & 0.269(1)   & 2.37 (2)   &     \\ 
        2.0 & eq.\ \rf{Q}                                                            & 2.93(1)   & 0.313(2)   &     &  $\sqrt{3}$   \\ 
        2.1 & eq.\ \rf{Q}                                                            & 3.63(1)   &  0.397(2)  &     &   $\sqrt{5}$  \\  
        2.2 & eq.\ \rf{Q}                                                            &  4.417(4)   & 0.498(1)   &     &  3   \\  
       2.25 & eq.\ \rf{Q}                                                            & 4.70(1)   & 0.541(2)   &     & $\sqrt{10}$    \\  
        2.3 &  eq.\ \rf{Q}                                                            &  4.812 
        \footnote{The actual constant derived from the fit to the linear portion of the data 
        was 4.77(1).  However, the correlator at $\b=2.3$ is extremely sensitive to value of $c_1$, and a small adjustment to 
        4.812 greatly improves the agreement with the Polyakov line correlator derived from lattice gauge theory.}
          & 0.563(2)   &     &  $\sqrt{13}$   \\  

        \hline
\end{tabular}
\caption{Constants defining the effective Polyakov line action \rf{SP1} for pure SU(2) lattice gauge theory on
a $16^3 \times 4$ lattice.} 
\label{tab1}
\end{center}
\end{table}

\section{\label{sec:conclude}Conclusions}

    We have calculated the effective Polyakov line actions corresponding to pure SU(2) lattice gauge theories on a
$16^3 \times 4$ lattice volumes in an interval of lattice couplings ranging from $\b=1.2$, which is deep in the strong-coupling
regime, up to $\b=2.3$, which is at the deconfinement transition.  At each coupling, the effective lattice action has a simple
bilinear form \rf{SP1}, although the range of the bilinear kernel $Q(\vx-\vy)$ varies from only nearest neighbor couplings,
in the strong-coupling regime, up to separations of $|\vx-\vy|=3.61$ lattice units at the deconfinement transition.  This extends
the work of ref.\ \cite{Greensite:2013yd}, where only the lattice coupling $\b=2.2$ was considered.  Our test for 
checking the validity of these effective actions is the comparison of Polyakov line correlators calculated in the effective theory and the corresponding lattice gauge theory.  In every case, the comparison works out quite well, down to correlator values
on the order of $10^{-5}$.  There have been other approaches to calculating the effective Polyakov line action, including
strong-coupling expansions \cite{Fromm:2011qi}, the Inverse Monte Carlo method \cite{Dittmann:2003qt,*Heinzl:2005xv},
and the Demon approach \cite{Velytsky:2008bh,*Wozar:2008nv}, resulting in effective actions of varying complexity, but we do not believe that these have yet demonstrated a comparable agreement in Polyakov line correlators at the larger $\b$ values,
at least not beyond two or three lattice spacings.

    So far we have not tried to calculate the effective action inside the deconfined regime.  Preliminary work suggests that
the range of the kernel in this region region is rather large in lattice units, which makes the effective action difficult to simulate, and also it may be necessary to introduce higher powers of $P_\vx$ in the potential of the effective theory.  We leave this problem for future investigation.

    Although we believe that deriving the effective Polyakov line action is of interest in itself, our ultimate goal is to apply our approach to the sign problem \cite{deForcrand:2010ys}.  There is no sign problem in SU(2), nor is there a sign problem in any pure gauge theory, so the next step in our program will be to apply the relative weights method to the SU(3) gauge group, first without and then with matter fields.  Our method only supplies the effective action at zero chemical potential.  However, as explained in 
ref.\ \cite{Greensite:2013yd}, the effective action corresponding to a lattice gauge theory at finite chemical potential may be obtained from the effective action at zero chemical potential by a simple change of variables.  If our method is successful for the SU(3) gauge group, then the strategy would be to apply one or more of the methods in refs.\ 
\cite{Gattringer:2011gq,*Mercado:2012ue,Fromm:2011qi,Aarts:2011zn,Greensite:2012xv}, which were developed to tackle the sign problem in Polyakov line actions,  to the problem of determining the phase structure of the theory.

\acknowledgments{J.G.'s research is supported in part by the
U.S.\ Department of Energy under Grant No.\ DE-FG03-92ER40711.  K.L.'s research is supported by STFC under the DiRAC framework. We are grateful for support from the HPCC Plymouth, where the numerical computations have been carried out.}     
   
\bibliography{pline}
 
\end{document}